\documentclass[aps,prd,twocolumn,groupedaddress,showpacs,nofootinbib,amssymb]{revtex4}
\usepackage[T1]{fontenc}
\usepackage[latin1]{inputenc}
\usepackage{graphicx}
\usepackage[english]{babel}
\usepackage{graphicx}
\usepackage{bm}
\usepackage{amsmath}
\usepackage{amssymb}
\usepackage{amsfonts}
\usepackage{epsfig}
\usepackage{colordvi}
\usepackage{color}

\begin{document}
\title{Quadrupolar gravitational radiation as a  test-bed for  $f(R)$-gravity}

\author{ Mariafelicia De Laurentis$^{1,2}$, Salvatore Capozziello$^{1,2}$}
\affiliation{\it $^1$Dipartimento di Scienze Fisiche, Università
di Napoli {}``Federico II'', 
$^2$INFN Sezione  di Napoli, \\Compl. Univ. di
Monte S. Angelo, Edificio G, Via Cinthia, I-80126, Napoli, Italy.}
\date{\today}

\begin{abstract}
The debate concerning the viability of 
 $f(R)$-gravity as a natural extension of General Relativity could be realistically addressed  by using  results coming from   binary pulsars like  PSR 1913+16.  To this end, we develop a quadrupolar approach to the gravitational radiation for a  class of analytic $f(R)$-models. We  show that experimental results  are compatible with a consistent range of $f(R)$-models. This  means that  $f(R)$-gravity is not ruled out by the observations and gravitational radiation (in strong field regime) could be a test-bed for such theories. 
\end{abstract}

\maketitle

\section{Introduction}
\label{uno}
The discovery of binary pulsars PSR 1913+16 by Hulse and Taylor in 1974 \cite{hulse} opened  a new testing ground for General Relativity (GR). In fact its
 continuous observation by Taylor and coworkers 
\cite{taylor,weisberg}, led to an impressively accurate tracking
of the orbital motion of the  binary system.  Before this discovery, the  only available testing ground for GR was the Solar System where the gravitational field is slowly varying 
and represents only a very small deformation of a flat space-time. As a consequence, Solar System tests can only prove the  weak-field limit of GR. By contrast, 
binary systems containing compact objects as neutron stars (NS) or black holes (BH) involve space-time domains where  the gravitational field is strong.
 Indeed, the  gravitational field on the surface ({\it i.e.} $\simeq 2 \, GM/c^2 R$) of a NS is of order $0.4$, which is close to the one of a BH $\simeq 2 \, GM/c^2 R = 1$) and much larger
 than the gravitational field on surfaces of Solar System bodies: $\simeq (2 \, GM/c^2 R)_{\odot} \sim 10^{-6}$, $(2 \, GM/c^2 R)_{\oplus} \sim 10^{-9}$. In addition, the high stability of 
 {\it pulsar clocks} has made it possible to monitor the dynamics of its orbital motion down to a precision allowing one to measure the small  $(\sim (v/c)^5)$ orbital effects linked to the propagation of the gravitational field at the velocity of light between the pulsar and its companion.
 The recent discoveries of the  double binary pulsars \cite{Burgay03,Lyne04} has renewed  the interest in the use of binary pulsars as extremely relevant test-beds of gravity theories.
This means that it is worth reconsidering in detail, {\it i.e.} at its foundation, the problem of motion also in relation to the problem of generation and detection of gravitational waves (GWs). In other words, the motion of sources could give further signatures to GWs and then it has to be carefully reconsidered.

The achieved sensitivity levels and theoretical developments are
leading toward a general picture of GW phenomena
that was not possible in the previous
pioneering era. Experimentally, several GW ground-based
laser interferometer detectors have been built
in the United States (LIGO) \cite{ligo}, Europe (VIRGO and
GEO) \cite{virgo,geo} and Japan (TAMA) \cite{tama}, and are now taking data
at designed sensitivities. A laser-interferometer space antenna
(LISA) \cite{lisa}  might fly within the next decade.
As results, we can  hope that the next decade will witness the direct detection of gravitational waves  opening  the fields of GW astronomy and  cosmology. Theoretical studies have been developed in parallel   to the experimental activity. In particular,  mechanisms  for the production of GWs, both in astrophysics and in cosmology.
Templates on binary inspiral  (see {\it e.g.} \cite{Buonanno1999,Cutler1993,Damour2003,Pan2004})
and robust search algorithms have been developed for GWs sources \cite{alggen1}. 
Furthermore conceptual and technical problems, related  to the production of GWs by self-gravitating systems (such as coalescing binaries) have not been fully solved.
This status of art suggests  to reconsider the problems of motion and  generation of GWs also with respect to alternative theories of gravity which seem realistic approaches to face several problems in astrophysics and cosmology.   In particular $f(R)$-gravity  seem a viable semi-classical  scheme to overcome shortcomings  related to infrared and ultraviolet behaviors of the gravitational field \cite{review}.  These theories   are based on corrections and enlargements of the Einstein GR.
Besides fundamental physics motivations, they have
acquired interest in cosmology due to the fact
that they "naturally" exhibit inflationary behaviors able to
overcome the shortcomings of Standard Cosmological
Model (based on GR). The related cosmological
models seem  realistic and, several times, capable
of matching with the observations \cite{la,kerner,tak,alle}.
From a genuine astrophysical viewpoint, these Extended Theories
of Gravity (ETGs) \cite{book} do not urgently require to find out candidates
for dark energy and dark matter at fundamental level
(till now they have not been detected!). The approach is very conservative
 taking into account only the "actually observed" ingredients
({\it i.e.} gravity, radiation and baryonic matter); it is in full
agreement with the early spirit of GR which
could not act in the same way at all scales (see \cite{book} for a comprehensive review). In fact, GR
 has been successfully probed in the weak-field
limit ({\it e.g.} Solar System experiments) and also in this case
there is room for alternative theories of gravity which are
not at all ruled out, as discussed in several recent studies
 \cite{will,lecian-solar,cap}. In particular, it is possible to show that several
$f(R)$-models could satisfy both cosmological and Solar System
tests \cite{cap2,hu}, could be constrained as the scalar-tensor
theories and could give rise to new effects capable of
explaining anomalies also at local scales  (see for example
\cite{bertolami} and  references therein).

In this paper we study the quadrupolar gravitational radiation in $f(R)$-gravity using the  "linearized theory". It consists in  expanding the field equations  around the flat Minkowski metric. The field equations then reduce to linear wave equations from which radiation can be calculated.
 GR predicts radiation that, at the lowest order, is proportional to the third derivative of the quadrupole momentum of the mass-energy distribution. It is a consequence of conservation equations that the first derivative of the monopole momentum and the second derivative of the dipole momentum are zero. This means that  the gravitational  radiation is first seen at the quadrupole term.  The  dipole effects depends on the difference of the self-gravitational binding energy per unit mass for two bodies and it is thus dependent also on the internal structures of the objects. When the objects are in circular orbits, the time variation of the scalar field at each object, due to the motion of the other,  is zero and the dipole contributions consequently drop out. Under these circumstances the dominant surviving terms are of quadrupole order.
In  $f(R)$-gravity the situation is different due to the presence of further degrees of freedom of  the gravitational field \cite{SCF,veltman,greci}. However, GR has to be fully recovered as soon as $f(R)\rightarrow R$. This "compatibility" with GR could be a test-bed for these  ETGs. Here we develop expressions for quadrupole gravitational radiation in $f(R)$-gravity using the weak field technique and apply these results, to  binary systems as, for example, the well known PSR 1913+16. In this way, it is straightforward to compare the GR-predictions 
with those of ETGs.
The outline of the paper is the following. In Sec. \ref{due}, we briefly introduce the weak field limit and field equations of $f(R)$-gravity.  Secs. \ref{tre} and \ref{quattro} are devoted to the  calculation of  the conservation laws. Finally the application to PSR 1913+16 is developed in Sec. \ref{cinque}. Conclusions are drawn in  Sec. \ref{sei}.

\section{ Field equations  and post-Mincowskian limit of $f(R)$-gravity}
\label{due}

 The post-Minkowskian limit of any theory of  gravity arises when the
regime of small field is considered without any prescription on
 the propagation  of the field. This case has to be
clearly distinguished with respect to the Newtonian limit which,
differently, requires both the small velocity and the weak field
approximations. Often, in literature, such a distinction is not
clearly remarked and several cases of pathological analysis can be
accounted. The post-Minkowskian limit of GR gives rise to
massless  gravitational waves. An analogous study can be pursued considering, instead of 
 the Hilbert-Einstein Lagrangian linear in the Ricci scalar $R$,  a general function $f(R)$ \cite{arturo}. The only assumption that we are going to do is that $f(R)$ is an analytic function. The gravitational action is then

\begin{equation}
{\cal A}=\int
d^4x\sqrt{-g}\biggl[f(R)+\mathcal{X}\mathcal{L}_m\biggr]\,,
\end{equation}
where ${\displaystyle \mathcal{X}=\frac{16\pi G}{c^4}}$ is the coupling,   $\mathcal{L}_m$ is the
standard matter Lagrangian and $g$ is the determinant of the metric\footnote{Here we indicates with "$,$" partial derivative and with " $;$" covariant derivative with regard to  $g_{\mu\nu}$; all Greek indices run from $0,...,3$ and Latin indices run from $1,...,3$; $g$ is the determinant.  }.
The field equations, in metric formalism, read\footnote{All considerations are developed here in metric formalism. From now on we assume physical  units $G=c= 1$.}

\begin{equation}\label{fe1}
f'(R)R_{\mu\nu}-\frac{1}{2}fg_{\mu\nu}-f'(R)_{;\mu\nu}+g_{\mu\nu}\Box_g
f'(R)=\frac{\mathcal{X}}{2}T_{\mu\nu}\,,
\end{equation}
\begin{equation}\label{TrHOEQ}
3\Box f'(R)+f'(R)R-2f(R)\,=\,\frac{\mathcal{X}}{2}T\,,
\end{equation}
with
${\displaystyle T_{\mu\nu}=\frac{-2}{\sqrt{-g}}\frac{\delta(\sqrt{-g}\mathcal{L}_m)}{\delta
g^{\mu\nu}}}$ the energy momentum tensor of matter ($T$ is the
trace), ${\displaystyle f'(R)=\frac{df(R)}{dR}}$ and
$\Box_g={{}_{;\sigma}}^{;\sigma}$. We adopt a $(+,-,-,-)$
signature, while the conventions for Ricci's tensor is
$R_{\mu\nu}={R^\sigma}_{\mu\sigma\nu}$ and 
${R^\alpha}_{\beta\mu\nu}=\Gamma^\alpha_{\beta\nu,\mu}+...$ for the Riemann tensor, where

\begin{equation}\label{chri}
\Gamma^\mu_{\alpha\beta}=\frac{1}{2}g^{\mu\sigma}(g_{\alpha\sigma,\beta}+g_{\beta\sigma,\alpha}-g_{\alpha\beta,\sigma})\,,
\end{equation}
are the Christoffel symbols  of the $g_{\mu\nu}$
metric. Actually, in order to perform a post-Minkowskian
limit of field equations, one has to perturb  Eqs.
(\ref{fe1})  on the Minkowski background
$\eta_{\mu\nu}$. In such a case  the invariant metric element  becomes

\begin{equation}\label{me} ds^2=g_{\sigma\tau}dx^\sigma
dx^\tau=(\eta_{\sigma\tau}+h_{\sigma\tau})dx^\sigma dx^\tau\,,
\end{equation}
with $h_{\mu\nu}$ small (${\mathcal O}(h)^2\ll 1$). 
We assume that the $f(R)$-Lagrangian is  analytic
 ({\it i.e.} Taylor expandable) in term of the Ricci scalar, which means that\footnote{for convenience we will use for the following calculations, $f$ instead of $f(R)$}
 
\begin{eqnarray}\label{sertay}
f(R)=\sum_{n}\frac{f^n(R_0)}{n!}(R-R_0)^n\simeq
f_0+f'_0R+\frac{1}{2}f''_0R^2+...\,.
\end{eqnarray}
The flat-Minkowski background is recovered for   $R=R_0\simeq 0$.  

Field equations (\ref{fe1}), at the first order of
approximation in term of the perturbation \cite{spher},
become\,:

\begin{equation}\label{fe2}
f_0'\biggl[R^{(1)}_{\mu\nu}-\frac{R^{(1)}}{2}\eta_{\mu\nu}\biggr]-f''_{0}\biggl[R^{(1)}_{,\mu\nu}-\eta_{\mu\nu}\Box
R^{(1)}\biggr]=\frac{\mathcal{X}}{2}T^{(0)}_{\mu\nu}
\end{equation}
where ${\displaystyle f'_0=\frac{df}{dR}\Bigl|_{R=0}}$,
${\displaystyle f''_0=\frac{d^2f}{dR^2}\Bigl|_{R=0}}$ and
$\Box={{}_{,\sigma}}^{,\sigma}$ that is now the standard d'Alembert operator of flat space-time. 
From the zero-order of Eqs.(\ref{fe1}),
one gets $f(0)=0$, while $T_{\mu\nu}$ is fixed at zero-order in
Eq.(\ref{fe2}) since, in this perturbation scheme, the first order on
Minkowski space has to be connected with the zero order of the
standard matter energy momentum tensor\footnote{This formalism
descends from the theoretical setting of Newtonian mechanics which
requires the appropriate scheme of approximation when obtained
from a more general relativistic theory. This scheme coincides
with a gravity theory analyzed at the first order of perturbation
in the curved spacetime metric.}. The explicit expressions of
the Ricci tensor and scalar, at the first order in the metric
perturbation, read

\begin{equation}\label{approx1}
\left\{\begin{array}{ll}R^{(1)}_{\mu\nu}=h^\sigma_{(\mu,\nu)\sigma}-\frac{1}{2}\Box
h_{\mu\nu}-\frac{1}{2}h_{,\mu\nu}\\\\
R^{(1)}={h_{\sigma\tau}}^{,\sigma\tau}-\Box h \end{array}\right.
\end{equation}
with $h={h^\sigma}_\sigma$. Eqs. (\ref{fe2}) can be written in a
more suitable form by introducing the constant
${\displaystyle \xi=-\frac{f''_0}{f'_0}}$, that is

\begin{eqnarray}\label{fe3}
&& h^\sigma_{(\mu,\nu)\sigma}-\frac{1}{2}\Box
h_{\mu\nu}-\frac{1}{2}h_{,\mu\nu}-\frac{1}{2}({h_{\sigma\tau}}^{,\sigma\tau}-\Box
h)\eta_{\mu\nu}\nonumber\\ &&+\xi(\partial^2_{\mu\nu}-\eta_{\mu\nu}\Box)({h_{\sigma\tau}}^{,\sigma\tau}-\Box
h)=\frac{\mathcal{X}}{2f'_0}T^{(0)}_{\mu\nu}\,.
\end{eqnarray}
By choosing the transformation
$\tilde{h}_{\mu\nu}=h_{\mu\nu}-\frac{h}{2}\eta_{\mu\nu}$ and the
gauge condition $\tilde{h}^{\mu\nu}_{\,\,\,\,\,\,\,,\mu}=0$, one
obtains that field equations  and the trace equation, respectively, read
\footnote{The gauge transformation is $h'_{\mu\nu}=h_{\mu\nu}-\zeta_{\mu,\nu}-\zeta_{\nu,\mu}$ when we
perform a coordinate transformation as $x'^\mu=x^\mu+\zeta^\mu$
with O($\zeta^2$)$\ll 1$. To obtain the gauge and the validity of
the field equations for both perturbation $h_{\mu\nu}$ and
$\tilde{h}_{\mu\nu}$, the $\zeta_\mu$ have to satisfy the harmonic
condition $\Box\zeta^\mu=0$.}

\begin{equation}\label{fe4}
\left\{\begin{array}{ll}\Box\tilde{h}_{\mu\nu}+\xi(\eta_{\mu\nu}\Box-\partial^2_{\mu\nu})\Box\tilde{h}=-\frac{\mathcal{X}}{f'_0}T^{(0)}_{\mu\nu}
\\\\
\Box\tilde{h}+3\xi\Box^2\tilde{h}=-\frac{\mathcal{X}}{f'_0}T^{(0)}\end{array}
\right.
\end{equation}
 It is worth  noticing that  solving the previous system of equations, we find  wavelike solutions with massless and massive contributions \cite{greci,arturo,SCF}.
The presence of the massive term is a feature emerging from the higher-order terms in $f(R)$-gravity. Specifically, it is related to the fact that  $f''_0\neq 0$, which is null in GR where $f(R)=R$. This means that massless states are a particular case among the gravitational theories that present also massive ones. A similar situation emerges also in the Newtonian limit:  the Newton potential is recovered only as the weak field limit of GR. In general, Yukawa-like corrections, and then characteristic interaction lengths, are present \cite{noi-newt}. The effective mass is $m^2= (3\xi)^{-1}=-\frac{f'_0}{3f''_0}$ and then $f''_0$ has to be negative in order to have physically defined states.   It is easy to see that massive modes are directly related to the non-trivial structure of the trace equation  Eq.(\ref{TrHOEQ}). In GR,  the Ricci scalar is univocally fixed being $R=0$ in vacuum and $R\propto \rho$ in presence of matter, where $\rho$ is the matter-energy density \cite{greci,arturo,SCF}.  The task  is now to 
evaluate the related energy-momentum tensors.
\section{Energy-Momentum Tensors}
\label{tre}
Let us assume that  the source $T_{\mu\nu}$ is localized in a finite region. Outside this region $T_{\mu\nu}=0$.

Then, as a consequence of Eqs.(\ref{approx1}) and gauge condition, we have
\begin{eqnarray}
R^{(1)}_{\mu\nu}=\Box h_{\mu\nu}=0\,,
\label{2.17}
\end{eqnarray}
outside the region.
There are several ways to define the energy-momentum tensor of the
gravitational field. One is to
consider $R_{\mu\nu}$ on the left-hand  side of Eq.(\ref{fe1})  consisting of a series of correction  terms in $R^{(N)}_{\mu\nu}$. 
In the development of Eq.(\ref{fe2}), $R_{\mu\nu}^{(1)}$ is on the left-hand side. The remaining higher order terms, which so far have been ignored, could be
brought to the right-hand side. If the source region gives rise to a flux of energy in the form
of GWs, it must be represented by these higher order terms. This is the geometric approach \cite{Maggiore}.
The other approach is to use the standard field theoretical methods. The geometric and the field theoretical approaches are complementary. Some aspects of GWs physics can be better understood from the former approach, some from the latter, and to study GWs from both vantage points results in a deeper overall understanding. 
We use the latter approach to calculate the
stress-energy tensor of the gravitational field. So one can extending the formalism to more general
theories and obtain this quantity by varying  the
gravitational Lagrangian. In GR, this quantity is a pseudo-tensor  and  is typically
referred to  as the Landau-Lifshitz energy-momentum tensor \cite{landau}.

In the case of $f(R)$-gravity, we have

\begin{eqnarray}
&& \delta\int d^4x\sqrt{-g}f(R)=\delta\int
d^4x\mathcal{L}(g_{\mu\nu},g_{\mu\nu,\rho},g_{\mu\nu,\rho\sigma})\approx\nonumber\\ &&\int
d^4x\biggl(\frac{
\partial\mathcal{L}}{\partial g_{\rho\sigma}}-\partial_\lambda\frac{\partial\mathcal{L}}{\partial
g_{\rho\sigma,\lambda}}+\partial^2_{\lambda\xi}\frac{\partial\mathcal{L}}{\partial
g_{\rho\sigma,\lambda\xi}}\biggr)\delta
g_{\rho\sigma}=\nonumber\\&&\doteq\int
d^4x\sqrt{-g}H^{\rho\sigma}\delta g_{\rho\sigma}=0\,.\nonumber\\
\end{eqnarray}

The Euler-Lagrange
equations are then

\begin{eqnarray}
\frac{\partial \mathcal{L}}{\partial
g_{\rho\sigma}}-\partial_\lambda\frac{\partial\mathcal{L}}{\partial
g_{\rho\sigma,\lambda}}+\partial^
2_{\lambda\xi}\frac{\partial\mathcal{L}}{\partial
g_{\rho\sigma,\lambda\xi}}=0,
\end{eqnarray}
which coincide with the field Eqs. (\ref{fe1}) in  vacuum.
Actually, even in the case of  more general theories,  it is possible to
define an energy-momentum tensor that 
 turns out to be defined as follows\,:
 
 \begin{eqnarray}
t^\lambda_\alpha&=&\frac{1}{\sqrt{-g}}\left[\left(\frac{\partial\mathcal{L}}{\partial
g_{\rho\sigma,\lambda}}-\partial_\xi\frac{\partial\mathcal
{L}}{\partial
g_{\rho\sigma,\lambda\xi}}\right)g_{\rho\sigma,\alpha}+\right.\nonumber\\ &&\left.+\frac{\partial\mathcal
{L}}{\partial
g_{\rho\sigma,\lambda\xi}}g_{\rho\sigma,\xi\alpha}-\delta^\lambda_\alpha\mathcal{L}\right]\,.
\end{eqnarray}

This quantity, together with the energy-momentum tensor of matter
$T_{\mu\nu}$, satisfies a conservation law as required by the Bianchi identities. In fact, in presence of matter, one has
${\displaystyle H_{\mu\nu}\,=\,\displaystyle\frac{\chi}{2}T_{\mu\nu}}$,  and then

\begin{eqnarray}
(\sqrt{-g}t^\lambda_\alpha)_{,\lambda}&=&-\sqrt{-g}H^{\rho\sigma}g_{\rho\sigma,\alpha}=\nonumber\\ &&=-\frac{\mathcal{X}}{2}\sqrt{-g}T^{\rho\sigma}
g_{\rho\sigma,\alpha}=-\mathcal{X}(\sqrt{-g}T^\lambda_\alpha)_{,\lambda}\,,\nonumber\\
\end{eqnarray}
and, as a consequence,

\begin{eqnarray}
[\sqrt{-g}(t^\lambda_\alpha+\mathcal{X}T^\lambda_\alpha)]_{,\lambda}=0\,,
\end{eqnarray}
that is  the conservation law given by the Bianchi identities.  We can now write  the
expression of the energy-momentum tensor $t^\lambda_\alpha$ in term of the gravity action $f(R)$ and its
 derivatives:

\begin{eqnarray}\label{tens-f(R)}
t^\lambda_\alpha&=&f'\biggl\{\biggl[\frac{\partial R}{\partial
g_{\rho\sigma,\lambda}}-\frac{1}{\sqrt{-g}}\partial_\xi\biggl(\sqrt{-g}
\frac{\partial R}{\partial
g_{\rho\sigma,\lambda\xi}}\biggr)\biggl]g_{\rho\sigma,\alpha}\biggl.\nonumber\\ &&+\biggl.\frac{\partial
R }{\partial
g_{\rho\sigma,\lambda\xi}}g_{\rho\sigma,\xi\alpha}\biggr\}-f''R_{,\xi}\frac{\partial
R }{\partial
g_{\rho\sigma,\lambda\xi}}g_{\rho\sigma,\alpha}-\delta^\lambda_\alpha\
f\,,\nonumber\\
\end{eqnarray}
It is worth noticing that $t^\lambda_\alpha$ is a non-covariant
quantity in GR while its  generalization, in fourth order gravity,
turns out to satisfy the covariance prescription of standard tensors (see also \cite{HOG}). On the other hand, such an
expression reduces to  the Landau-Lifshitz
energy-momentum tensor of GR as soon as  $f(R)\,=\,R$, that is
\begin{eqnarray}
{t^\lambda_\alpha}_{|_{\text{GR}}}=\frac{1}{\sqrt{-g}}\biggl(\frac{\partial\mathcal{L}_{\text{GR}}}{\partial g_{\rho\sigma,\lambda}}g_{\rho\sigma
,\alpha}-\delta^\lambda_\alpha\mathcal{L}_{\text{GR}}\biggr)\,,
\end{eqnarray}
where the GR Lagrangian has been considered in its effective form,
{\it i.e.} the symmetric part of the Ricci tensor, which effectively
 leads to the 
equations of motion, that is 
\begin{equation}
\mathcal{L}_{\text{GR}}=\sqrt{-g}g^{\mu\nu}(\Gamma^\rho_{\mu\sigma}\Gamma^\sigma_{\rho\nu}-\Gamma^\sigma_{\mu\nu}\Gamma^\rho_{\sigma\rho})\,.
\end{equation}
It is important to stress that the definition of the
energy-momentum tensor in GR and in $f(R)$-gravity are  different.
This discrepancy is due to the presence, in the second case,
of higher than second order differential terms  that  cannot be discarded  by means of a boundary
integration as it is done in GR. We have noticed  that the
effective Lagrangian of GR turns out to be the symmetric part of
the Ricci scalar since the second order terms, present in the
definition of $R$ , can be removed by means of integration by parts.

On the other hand, 
an analytic $f(R)$-Lagrangian can be recast,  at linear order,
as $f\sim f'_0R+\mathcal{F}(R)$, where the function
$\mathcal{F}$ satisfies the condition: $\lim_{R\rightarrow
0}\mathcal{F}\rightarrow R^2$. As a consequence, one can rewrite
the explicit expression of $t^\lambda_\alpha$ as\,:

\begin{eqnarray}\label{tensorf(R).1}
t^\lambda_\alpha&=&f'_0{t^\lambda_\alpha}_{|_{\text{GR}}}+\nonumber\\&&+\mathcal{F}'\biggl\{\biggl[\frac{\partial
R}{\partial g_{\rho
\sigma,\lambda}}-\frac{1}{\sqrt{-g}}\partial_\xi\biggl(\sqrt{-g}\frac{\partial
R}{\partial
g_{\rho\sigma,\lambda\xi}}\biggr)\biggl]g_{\rho\sigma,\alpha}\biggl.\nonumber\\&&+\biggl.\frac{\partial
R }{\partial
g_{\rho\sigma,\lambda\xi}}g_{\rho\sigma,\xi\alpha}\biggr\}-\mathcal{F}''R_{,\xi}\frac{\partial
R }{\partial
g_{\rho\sigma,\lambda\xi}}g_{\rho\sigma,\alpha}-\delta^\lambda_\alpha
\mathcal{F}.\nonumber\\
\end{eqnarray}
The general expression of the Ricci scalar, obtained by splitting
its linear ($R^*$) and quadratic ($\bar{R}$) parts  once a
perturbed metric (\ref{me}) is considered, is
\begin{eqnarray}\label{defRicciscalar}
R&=&g^{\mu\nu}(\Gamma^\rho_{\mu\nu,\rho}-\Gamma^\rho_{\mu\rho,\nu})+g^{\mu\nu}(\Gamma^{\rho}_{\sigma\rho}\Gamma^
{\sigma}_{\mu\nu}-\Gamma^{\sigma}_{\rho\mu}\Gamma^{\rho}
_{\nu\sigma})=\nonumber\\ && = R^*+\bar{R}\,,
\end{eqnarray}
(notice that $\mathcal{L}_{\text{GR}}=-\sqrt{-g}\bar{R}$).
In the case of GR ${t^\lambda_\alpha}_{|_{\text{GR}}}$,
the Landau-Lifshitz tensor presents a first non-vanishing term at order
 $h^2$. A similar result can be obtained in the case of $f(R)$-gravity.  In fact, taking into account Eq.(\ref{tensorf(R).1}), 
 one obtains that, at the lower order,
$t^\lambda_\alpha$ reads\,:
\begin{eqnarray}\label{tensorf(R).2}
t^\lambda_\alpha&\sim&{t^\lambda_\alpha}_{|h^2}=f'_0{t^\lambda_\alpha}_{|_{\text{GR}}}+\nonumber\\ &&+ f''_0R^*\biggl[\biggl(-
\partial_\xi\frac{\partial R^*}{\partial
g_{\rho\sigma,\lambda\xi}}\biggr)g_{\rho\sigma,\alpha}+\frac{\partial
R^*}{\partial
g_{\rho\sigma,\lambda\xi}}g_{\rho\sigma,\xi\alpha}\biggr]-\nonumber\\ && +f''_0R^*_{,\xi}\frac{\partial
R^*}{\partial
g_{\rho\sigma,\lambda\xi}}g_{\rho\sigma,\alpha}-\frac{1}{2}f''_0\delta^\lambda_\alpha
{R^*}^2=\nonumber\\
&=&f'_0{t^\lambda_\alpha}_{|_{\text{GR}}}+f''_0\biggl[R^*\biggl(\frac{\partial
R^*}{\partial
g_{\rho\sigma,\lambda\xi}}g_{\rho\sigma,\xi\alpha}-\frac{1}{2}R^*\delta^\lambda_\alpha\biggr)-\biggr.\nonumber\\&&\biggl.+\partial_\xi\biggl(R^*\frac{\partial
R^*}{\partial
g_{\rho\sigma,\lambda\xi}}\biggr)g_{\rho\sigma,\alpha}\biggr]\,.
\end{eqnarray}
Considering the perturbed metric (\ref{me}), we have $R^*\sim R^{(1)}$,
where $R^{(1)}$ is defined as in (\ref{approx1}). In terms of $h$ and $\eta$, we get

\begin{eqnarray}
\left\{\begin{array}{ll}\frac{\partial R^*}{\partial
g_{\rho\sigma,\lambda\xi}}\sim\frac{\partial R^{(1)}}{\partial
h_{\rho\sigma,\lambda\xi}}=\eta^{\rho\lambda}\eta^{\sigma\xi}-\eta^{\lambda\xi}\eta^{\rho\sigma}\\\\\frac{\partial
R^*}{\partial
g_{\rho\sigma,\lambda\xi}}g_{\rho\sigma,\xi\alpha}\sim
h^{\lambda\xi}_{\,\,\,\,\,\,,\xi\alpha}-h^{,\lambda}_{\,\,\,\,\,\alpha}
\end{array}\right.\,.
\end{eqnarray}
Clearly, the first significant term in Eq. (\ref{tensorf(R).2}) is of
second order in the perturbation expansion. We can now write  the
expression of the energy-momentum tensor explicitly in term of the
perturbation $h$; it is 

\begin{eqnarray}
t^\lambda_\alpha&\sim&f'_0{t^\lambda_\alpha}_{|_{\text{GR}}}+f''_0\{(h^{\rho\sigma}_{\,\,\,\,\,\,\,,\rho\sigma}-\Box
h)\left [h^{\lambda\xi}_{\,
\,\,\,\,\,\,,\xi\alpha}-h^{,\lambda}_{\,\,\,\,\,\,\,\alpha}-\right.\nonumber\\&&\left.+\frac{1}{2}\delta^\lambda_\alpha(h^{\rho\sigma}_{\,\,\,\,\,\,\,,
\rho\sigma}-\Box
h)\right]-h^{\rho\sigma}_{\,\,\,\,\,\,\,,\rho\sigma\xi}h^{\lambda\xi}_{\,\,\,\,\,\,\,,\alpha}+\nonumber\\ &&+h^{\rho\sigma\,\,\,\,\,\,\,\,\,\,\lambda}_
{\,\,\,\,\,\,\,,\rho\sigma}h_{,\alpha}+h^{\lambda\xi}_{\,\,\,\,\,\,\,,\alpha}\Box
h_{,\xi}-\Box h^{,\lambda}h_{,\alpha}\}\,.\nonumber\\
\label{t}
\end{eqnarray}
Considering the tilded perturbation metric $\tilde{h}_{\mu\nu}$, the more compact form

\begin{eqnarray}
{t^\lambda_\alpha}_{|_f}&=&\biggl[\frac{1}{4}\tilde{h}^{,\lambda}_{\,\,\,\,\alpha}\Box\tilde{h}-\frac{1}{4}\tilde{h}_{,\alpha}
\Box\tilde{h}^{,\lambda}-\frac{1}{2}\tilde{h}^{\lambda}_{\,\,\,\,\,\sigma,\alpha}\Box\tilde{h}^{,\sigma}-\Biggr.\nonumber\\ && \Biggl.+\frac{1}{8}(\Box\tilde{h})^2\delta^
\lambda_\alpha\biggr]\nonumber\,,
\end{eqnarray}
is achieved.

As matter of facts, the energy-momentum tensor of the gravitational
field, which expresses the energy transport  during
the propagation, has a natural generalization in the case of
$f(R)$-gravity. We have adopted here the
Landau-Lifshitz definition but other approaches can be taken into account  \cite{multamaki}. The general definition of
${t^\lambda_\alpha}$, obtained above, consists of a
sum of  a GR contribution plus a term coming from 
$f(R)$-gravity\,:

\begin{eqnarray}
t^\lambda_\alpha=f'_0{t^\lambda_\alpha}_{|_{\text{GR}}}+f''_0{t^\lambda_\alpha}_{|_f}\,.
\end{eqnarray}
However,  as soon as  $f(R)=R$,  we obtains
$t^\lambda_\alpha={t^\lambda_\alpha}_{|_{\text{GR}}}$. 
As a final remark, it is worth noticing  that massive modes of gravitational field come out from  ${{t^\lambda_\alpha}}_{|_f}$ since $\Box\tilde{h}$ can be considered an effective scalar field moving in a potential: $t^\lambda_\alpha$, in this case,  represents a transport tensor.

The expression for the gravitational tensor $t^\lambda_\alpha$ 
can be
simplified  by doing approximations valid far from the
source region. Far from the source $h_{\mu\nu}$ will be, functions of a single scalar variable $t'$
\begin{equation}
t'=t-r\,,
\label{2.20}
\end{equation}
where
\begin{equation}
r^2=x_ix^i\,.
\label{2.21}
\end{equation}
Such a scalar can be constructed from the vector $x^\mu$ by forming
\begin{equation}
t'=k_\lambda x^\lambda\,,
\label{2.22}
\end{equation}
with
\begin{equation}
k_0\equiv-k^0\equiv 1\,,\qquad k_i\equiv-{\hat x}_i\,,
\label{2.23}
\end{equation}

\begin{equation}
{\hat x}_i\equiv\frac{x^i}{r}\,.
\label{2.24}
\end{equation}
 In the far field $k_\lambda$ can be considered  as a constant vector, over a  small region. 
That is,  $h_{\mu\nu}$ will be almost plane. Any $\displaystyle{\frac{1}{r}}$ variation or change of the
unit vector ${\hat x}$ over points in the region can be made arbitrarily small by choosing a
region far from the source \cite{Maggiore}. 

The functional dependency of solutions will be on the  $t'=t-r$. This fact can be done by expressing all partials of  $h_{\mu\nu}$ as

\begin{equation}
h_{\mu\nu,\sigma}=\frac{\partial t'}{\partial x^\sigma}\frac{dh_{\mu\nu}}{dt'}=k_{\lambda}\delta^{\lambda}_{\sigma}{\dot h}_{\mu\nu}=k_\sigma{\dot h}_{\mu\nu}\,,
\label{2.25}
\end{equation}
where
\begin{equation}
h_{\mu\nu}=h_{\mu\nu}\left(k_\lambda x^\lambda\right)=h_{\mu\nu}(t')\,,
\label{2.27}
\end{equation}

here the dot indicate the derivative with respect to the time and $\displaystyle{\frac{\partial x^\lambda}{\partial x^\sigma}=\delta^{\lambda}_{\sigma}}$.
Since $T_{\mu\nu}=0$ outside the source region, 
\begin{equation}
\Box h_{\mu\nu}=0\,,
\label{2.31}
\end{equation}
in the far field \cite{greci,SCF}. If Eq.(\ref{2.25}) is used in the first of these, we find
\begin{equation}
\Box h_{\mu\nu}=h_{\mu\nu,\rho,}\,^{\rho}=\left(k_\rho {\dot h}_{\mu\nu}\right),^{\rho}=k_\rho k^\rho {\ddot h}_{\mu\nu}\,,
\end{equation}
implying that
\begin{equation}
k_\rho k^\rho=0\,.
\label{2.32}
\end{equation}
Therefore, from Eq.(\ref{t}), the energy-momentum tensor associated
with the tensor part of the gravitational field is
\begin{eqnarray}
t^\lambda_\alpha&=&f'_0 \left(k^\lambda k_\alpha {\dot h}^{\rho\sigma}  {\dot h}_{\rho\sigma}\right)+f''_0\left(k_\rho k_\sigma{\ddot h}^{\rho\sigma}k_\xi k_\alpha{\ddot h}^{\lambda\xi}-\right.\nonumber\\ &&\left.+k_\rho k_\sigma {\ddot h}^{\rho\sigma}k^\lambda k_\alpha {\ddot h}-\frac{1}{2}k_\rho k_\sigma {\ddot h}^{\rho\sigma} \delta^\lambda_\alpha k_\rho k_\sigma {\ddot h}^{\rho\sigma} +\right.\nonumber\\ &&\left.+\frac{1}{2}k_\rho k_\sigma {\ddot h}^{\rho\sigma}\delta^\lambda_\alpha \Box h-k_\xi k_\alpha {\ddot h}^{\lambda\xi}\Box h+k^\lambda k_\alpha {\ddot h}\Box h+\right.\nonumber\\ && \left.+\frac{1}{2}\delta^\lambda_\alpha k_\rho k_\sigma {\ddot h}^{\rho\sigma} \Box h-\frac{1}{2}\delta^\lambda_\alpha (\Box h)^2-k_\rho k_\sigma k_\xi {\dddot h}^{\rho\sigma} k_\alpha {\dot h}^\lambda\xi+\right.\nonumber\\ &&\left.+k_\rho k_\sigma {\dddot h}^{\rho\sigma}k^\lambda k_\alpha {\dot h}+k_\alpha {\dot h}^\lambda\xi \Box h_{,\xi}-\Box h^{,\lambda} k_\alpha {\dot h}\right)\,.\nonumber\\
\label{tderi}
\end{eqnarray}
Now remember that
\begin{equation}
{\dot h}=\eta_{\xi\lambda}{\dot h}^{\lambda\xi}\,,\qquad {\ddot h}=\eta_{\xi\lambda}{\ddot h}^{\lambda\xi}\,,
\end{equation}
and 
\begin{equation}
k^\lambda \eta_{\xi\lambda}=k_\xi\,,
\end{equation}
and then
\begin{equation}
k^\lambda k_\alpha {\ddot h}= k^\lambda k_\alpha \eta_{\xi\lambda}{\ddot h}^{\lambda\xi}=k_\xi k_\alpha {\ddot h}^{\lambda\xi}\,,
\end{equation}
we can further simplify $t^\lambda_\alpha$ in the following way
\begin{eqnarray}
t^\lambda_\alpha&=&f'_0 \left(k^\lambda k_\alpha {\dot h}^{\rho\sigma}  {\dot h}_{\rho\sigma}\right)+f''_0\left( k_\rho k_\sigma {\ddot h}^\rho\sigma k_\xi k_\alpha {\ddot h}^{\lambda\xi}+\right.\nonumber\\ && \left. -k_\rho k_\sigma {\ddot h}^{\rho\sigma}k^\lambda k_\alpha {\ddot h}-\frac{1}{2}{\ddot h}^{\rho\sigma}\delta^\lambda_\alpha k_\rho k_\sigma {\ddot h}^\rho\sigma-\right.\nonumber\\ && \left.+ k_\rho k_\sigma k_\xi {\dddot h}^{\rho\sigma}k_\alpha {\dot h}^{\lambda\xi}+ k_\rho k_\sigma {\dddot h}^{\rho\sigma} k^\lambda k_\alpha {\dot h}\right)\,,
\label{tderi1}
\end{eqnarray}
we notice that the sixth and fifth terms of above equation are equal because
\begin{eqnarray}
k_\rho k_\sigma k^\lambda k_\alpha {\dddot h}^{\rho\sigma} {\dot h}&=& k_\rho k_\sigma k^\lambda {\dddot h}^{\rho\sigma} \eta_{\xi\lambda}{\dot h}^{\lambda\xi}=\nonumber\\ &&= k_\rho k_\sigma k_\xi {\dddot h}^{\rho\sigma} {\dot h}^{\rho\xi}\,,
\end{eqnarray}
the third and fourth are the same, and then $t^\lambda_\alpha$ reduces to
 \begin{eqnarray}
t^\lambda_\alpha&=&f'_0 \left(k^\lambda k_\alpha {\dot h}^{\rho\sigma}  {\dot h}_{\rho\sigma}\right)-\frac{1}{2}f''_0 \left(k_\rho k_\sigma {\ddot h}^{\rho\sigma}\delta^\lambda_\alpha k_\rho k_\sigma {\ddot h}^{\rho\sigma}\right)=\nonumber\\&&=
f'_0 \left(k^\lambda k_\alpha {\dot h}^{\rho\sigma}  {\dot h}_{\rho\sigma}\right)-\nonumber\\&&+\frac{1}{2}f''_0\left(k_\rho k_\sigma {\ddot h}^{\rho\sigma}\eta^{\lambda\xi} \eta_{\xi\alpha} k_\rho k_\sigma {\ddot h}^{\rho\sigma}\right)\,,\nonumber\\
\label{tderi2}
\end{eqnarray}
finally the energy momentum tensor assume the following form
\begin{eqnarray}
t^\lambda_\alpha&=&\underbrace{f'_0 k^\lambda k_\alpha \left({\dot h}^{\rho\sigma}  {\dot h}_{\rho\sigma}\right)}_{GR}-\underbrace{\frac{1}{2}f''_0 \delta^{\lambda}_\alpha\left(k_\rho k_\sigma {\ddot h}^{\rho\sigma}\right)^2}_{f(R)}\,.
\label{tderi3}
\end{eqnarray}
To be more precise, the first term, depending on the choice of the constant $f'_0$, is the standard GR term,  the second is the $f(R)$ contribution. It is worth noticing that the order of derivative is increased of two degrees consistently to the fact that $f(R)$-gravity is of fourth-order in the metric approach.

Now we could compute the
instantaneous $\displaystyle{\frac{dE}{dt}}$ using the Eq.(\ref{tderi3}) as a basis.
The effect on a binary system is is more evident if we consider
 the average flux of energy away from the system.
Suppose that the $h_{\mu\nu}$ waves  can be represented by a discrete spectral representation.
 The periodicity $T$  will be proportional to
the inverse of the difference of the pair of frequency components in the wave.
Therefore, we must to evaluate the average of $\displaystyle{\frac{dE}{dt}}$ over an interval
equal to or greater than $T$ \cite{landau,Maggiore}.
The instantaneous flux of energy through a surface of area $r^2 d\Omega$ in the direction ${\hat x}$
\begin{equation}
\frac{dE}{dt}=r^2d\Omega {\hat x}^i t^{0i}\,,
\label{2.44}
\end{equation}
and the average flux is
\begin{equation}
\left\langle \frac{dE}{dt}\right\rangle=r^2d\Omega {\hat x}^i \langle t^{0i}\rangle\,,
\label{2.45}
\end{equation}
and then Eq. {\ref{tderi3}} becomes
\begin{eqnarray}
\left\langle t^\lambda_\alpha\right\rangle&=&\left\langle f'_0 k^\lambda k_\alpha \left({\dot h}^{\rho\sigma}  {\dot h}_{\rho\sigma}\right)-\frac{1}{2}f''_0 \delta^{\lambda}_\alpha\left(k_\rho k_\sigma {\ddot h}^{\rho\sigma}\right)^2 \right\rangle\,.\nonumber\\
\label{tmedia}
\end{eqnarray}

Finally, we re-write $\left\langle t^\lambda_\alpha\right\rangle$ in terms of a function $J_{\mu\nu}$ defined to be
\begin{eqnarray}
J_{\mu\nu}(\vec{x},t)\simeq 4\int d^3 {\vec x}' \frac{T_{\mu\nu}(\vec x', t-|\vec x'-\vec x|)}{|\vec x'-\vec x|}\,,
\label{2.50}
\end{eqnarray}
noting that
\begin{eqnarray}
h_{\mu\nu}(\vec{x},t)=J_{\mu\nu}(\vec{x},t)\,,
\label{2.51}
\end{eqnarray}
and consequently
\begin{eqnarray}
{\dot h}^{\rho\sigma}{\dot h}_{\rho\sigma}={\dot J}^{\rho\sigma}{\dot J}_{\rho\sigma}\qquad {\ddot h}^{\rho\sigma}{\ddot h}_{\rho\sigma}={\ddot J}^{\rho\sigma}{\ddot J}_{\rho\sigma} \label{2.52}
\end{eqnarray}

to give
\begin{eqnarray}
\left\langle t^\lambda_\alpha\right\rangle&=& \left\langle f'_0 k^\lambda k_\alpha  {\dot J}^{\rho\sigma}{\dot J}_{\rho\sigma}- \frac{1}{2}f''_0 \delta^{\lambda}_\alpha \left(k_\rho k_\sigma\right)^2     {\ddot J}^{\rho\sigma}{\ddot J}_{\rho\sigma} \right\rangle\,.\nonumber\\
\label{2.55}
\end{eqnarray}

\section{Momenta and Conservation Laws}
\label{quattro}

Let us  now analyze the radiation in terms of multipoles, that means to expand $J_{\mu\nu}$ in a Taylor series about $t'=t-r$. That is,
\begin{eqnarray}
&&J^{\mu\nu}(\vec{x},t)=\frac{4}{r}\left[\int d^3 {\vec x}' T^{\mu\nu}(\vec{x}',t')+\right. \nonumber\\ &&
\left. +{\hat x}\int d^3 {\vec x}'{\vec x}'\frac{\partial T^{\mu\nu}(\vec{x},t)}{\partial t'}+\right.\nonumber\\&&+\left.\frac{1}{2}\int d^3 {\vec x}'({\hat x}\cdot {\vec x}')^2 \frac{\partial^2 T^{\mu\nu}(\vec{x},t)}{\partial t'}^2\right]\,,
\label{2.56}
\end{eqnarray}
where we have used
\begin{eqnarray}
|\vec x'-\vec x|^{-1}\simeq\frac{1}{r}\,,
\end{eqnarray}
and
\begin{eqnarray}
|\vec x'-\vec x|\simeq r-{\hat x}\cdot {\vec x}'\,,
\end{eqnarray}
for $r>>|{\vec x}'|$.
Let us define the following moments of the mass-energy distribution:
\begin{eqnarray}
M(t)\simeq\int d^3 {\vec x}\, T^{00}(\vec{x},t)\,,
\label{2.57}
\end{eqnarray}
\begin{eqnarray}
D^{k}(t)\simeq\int d^3 {\vec x}\, x^k T^{00}(\vec{x},t)\,,
\label{2.58}
\end{eqnarray}
\begin{eqnarray}
Q^{ij}(t)\simeq\int d^3 {\vec x}\, x^i x^j T^{00}(\vec{x},t)\,.
\label{2.59}
\end{eqnarray}

The conservation law becomes, in the weak field limit,
\begin{eqnarray}
T^{\mu\nu}\,,_{\nu}=0\,,
\label{2.60}
\end{eqnarray}
and implies the relations \cite{Maggiore, gravitation}
\begin{eqnarray}
\int d^3 {\vec x}\, T^{jk}(\vec{x},t)&=&\frac{1}{2}\frac{\partial^2}{\partial t^2}\int d^3 {\vec x}\, x^j x^k T^{00}(\vec{x},t)=
\nonumber\\&&=\frac{1}{2}{\ddot Q}^{jk}(t)\,,
\label{2.61}
\end{eqnarray}
\begin{eqnarray}
\int d^3 {\vec x}\, T^{0k}(\vec{x},t)&=&\frac{\partial}{\partial t}\int d^3 {\vec x}\,  x^k T^{00}(\vec{x},t)=
\nonumber\\&&={\dot D}^{k}(t)\,,
\label{2.62}
\end{eqnarray}
\begin{eqnarray}
\frac{\partial}{\partial t}\int d^3 {\vec x}\, x^kT^{j0}(\vec{x},t)&=&\int d^3 {\vec x}\,  T^{jk}(\vec{x},t)=
\nonumber\\&&=\frac{1}{2}{\ddot Q}^{jk}(t)\,,
\label{2.63}
\end{eqnarray}
We use Eq.(\ref{2.56}) to write $J^{00}$, $J^{0i}$ and $J^{ij}$ in terms of the momenta Eqs.(\ref{2.57})-(\ref{2.59}).
First, from Eq.(\ref{2.56}) we have
\begin{eqnarray}
&& J^{00}(\vec{x},t)=4\frac{1}{r}\left[ \int d^3 {\vec x}'\,T^{00}(\vec{x}',t')+\right.\nonumber\\&&\left.+ 
{\hat x}_i \frac{\partial}{\partial t'} \int d^3 {\vec x}'\, x'^{i}T^{00}(\vec{x}',t' )+\right.\nonumber\\&&
\left.+ \frac{1}{2}{\hat x}_i{\hat x}_j \frac{\partial^2}{\partial t^2}\int d^3 {\vec x}'\, x'^{i} x'^{j} T^{00}(\vec{x}',t')+...\right]\,,\nonumber\\
\label{2.64}
\end{eqnarray}
For $J^{00}$ ,  it is easy to obtain
\begin{eqnarray}
J^{00}(\vec{x},t)=\frac{4}{f'_0}\frac{1}{r}\left[M(t')+{\hat x}_i {\dot D}^i(t')+\frac{1}{2}{\hat x}_i {\hat x}_j {\ddot Q}^{ij}(t')\right]\,.\nonumber\\
\label{2.65}
\end{eqnarray}
For $J^{0i}$  we need only two terms of the expansion Eq.(\ref{2.56})  in order to include terms up to
the second momentum
\begin{eqnarray}
J^{0i}(\vec{x},t)&=& 4\frac{1}{r}\left[ \int d^3 {\vec x}'\,T^{0i}(\vec{x}',t')+\right.\nonumber\\&&
\left.+{\hat x}_k \frac{\partial}{\partial t'} \int d^3 {\vec x}'\, x'^{k}T^{0i}(\vec{x}',t' )\right]\,,
\label{2.66}
\end{eqnarray}
Eq.(\ref{2.62}) and Eq.(\ref{2.63}) then give
\begin{eqnarray}
J^{0i}(\vec{x},t)=4\frac{1}{r}\left[{\dot D}^i(t')+\frac{1}{2}{\hat x}_k {\ddot Q}^{ik}(t')\right]\,.
\label{2.67}
\end{eqnarray}
For $J^{ij}$ only one term of Eq.(\ref{2.56}) is required, being
\begin{eqnarray}
J^{ij}(\vec{x},t)=2\frac{1}{r}{\ddot Q}^{ij}(t')\,.
\label{2.68}
\end{eqnarray}
The conservation law also implies that
\begin{eqnarray}
{\dot M}=0\,,\qquad {\ddot D}^k=0\,.
\label{2.69}
\end{eqnarray}

Furthermore, from Eq.(\ref{2.65}), Eq.(\ref{2.67}), and Eq.(\ref{2.68}) we have
\begin{eqnarray}
{\dot J}^{00}=2\frac{1}{r} {\hat x}_i{\hat x}_j{ \dddot Q}^{ij}\,,
\label{2.70}
\end{eqnarray}

\begin{eqnarray}
{\dot J}^{0i}=2\frac{1}{r}{\hat x}_k { \dddot Q}^{ik}\,,
\label{2.71}
\end{eqnarray}

\begin{eqnarray}
{\dot J}^{ij}=2\frac{1}{r} { \dddot Q}^{ij}\,,
\label{2.72}
\end{eqnarray}
and consequently

\begin{eqnarray}
{\ddot J}^{00}=2\frac{1}{r} {\hat x}_i{\hat x}_j{ \ddddot Q}^{ij}\,,
\label{270}
\end{eqnarray}

\begin{eqnarray}
{\ddot J}^{0i}=2\frac{1}{r}{\hat x}_k { \ddddot Q}^{ik}\,,
\label{271}
\end{eqnarray}

\begin{eqnarray}
{\ddot J}^{ij}=2\frac{1}{r} { \ddddot Q}^{ij}\,,
\label{272}
\end{eqnarray}

In order to evaluate Eq.(\ref{2.55}), we require that
\begin{eqnarray}
{\dot J}^{\rho\sigma}{\dot J}_{\rho\sigma}={\dot J}^{00}{\dot J}_{00}+2 {\dot J}^{0i}{\dot J}_{0i}+{\dot J}^{ij}{\dot J}_{ij}\,.
\label{2.73}
\end{eqnarray}
and 
\begin{eqnarray}
{\ddot J}^{\rho\sigma}{\ddot J}_{\rho\sigma}={\ddot J}^{00}{\ddot J}_{00}+2 {\ddot J}^{0i}{\ddot J}_{0i}+{\ddot J}^{ij}{\ddot J}_{ij}\,.
\label{273}
\end{eqnarray}

Pluggins Eqs.(\ref{2.70})-(\ref{272}) into Eq.(\ref{2.73}) and (\ref{273}), we get

\begin{eqnarray}
{\dot J}^{\rho\sigma}{\dot J}_{\rho\sigma}&=&\frac{4}{r^2}\left[\left(  {\hat x}_i{\hat x}_j{ \dddot Q}^{ij}\right)^2-\right.\nonumber\\&&\left.- 2 \left( {\hat x}_k{ \dddot Q}^{ik}\right)\left( {\hat x}_j{ \dddot Q}^{ij}\right)+\left({ \dddot Q}^{ij}{ \dddot Q}_{ij}\right)\right] \,.\nonumber\\
\label{2.75}
\end{eqnarray}

In completely analogous way, we find
\begin{eqnarray}
{\ddot J}^{\rho\sigma}{\ddot J}_{\rho\sigma}&=&\frac{4}{r^2}\left[\left(  {\hat x}_i{\hat x}_j{ \ddddot Q}^{ij}\right)^2-\right.\nonumber\\&&\left.- 2 \left( {\hat x}_k{ \ddddot Q}^{ik}\right)\left( {\hat x}_j{ \ddddot Q}^{ij}\right)+\left({ \ddddot Q}^{ij}{ \ddddot Q}_{ij}\right)\right] \,.\nonumber\\
\label{275}
\end{eqnarray}

When Eq.(\ref{2.75}) and Eq.(\ref{275}) are put into Eq.(\ref{2.55}), we 
find

\begin{eqnarray}
\left\langle t^\lambda_\alpha\right\rangle&=& \left\langle f'_0 k^\lambda k_\alpha \frac{4}{r^2}\left[\left(  {\hat x}_i{\hat x}_j{ \dddot Q}^{ij}\right)^2- 2 \left( {\hat x}_k{ \dddot Q}^{ik}\right)\left( {\hat x}_j{ \dddot Q}^{ij}\right)+\right.\right.\nonumber\\ &&\left.\left.+\left({ \dddot Q}^{ij}{ \dddot Q}_{ij}\right)\right]
- f''_0 \delta^{\lambda}_\alpha \left(k_\rho k_\sigma\right)^2  \frac{2}{r^2}\left[\left(  {\hat x}_i{\hat x}_j{ \ddddot Q}^{ij}\right)^2+\right.\right.\nonumber\\&&\left.\left.- 2 \left( {\hat x}_k{ \ddddot Q}^{ik}\right)\left( {\hat x}_j{ \ddddot Q}^{ij}\right)+\left({ \ddddot Q}^{ij}{ \ddddot Q}_{ij}\right)\right]  
 \right\rangle\,.\nonumber\\
\label{2.79}
\end{eqnarray}

Using the result in Eq.(\ref{2.45}) and integrating over all directions in order to
compute the total average flux of energy due to the tensor wave,

\begin{equation}
\left\langle \frac{dE}{dt}\right\rangle_{(total)}=r^2\int d \Omega {\hat x}^i
\langle t^{0i}\rangle\,.\label{2.80}\end{equation}
Note that
\begin{equation}
{\hat x}^\alpha\langle t^{0i}\rangle={\hat x}^i k^0 k^i [...]={\hat x}^i(-1)(-{\hat x}^i)[...]=[...]\, ,
\end{equation}
which simplify the evaluation of Eq.(\ref{2.80}). Integration over all direction is accomplished readily with the help of
\begin{equation}
\int d \Omega {\hat x}^i {\hat x}^j=\frac{4\pi}{3}\delta_{ij}\,,\label{2.81}
\end{equation}
and
\begin{equation}
\int d \Omega {\hat x}^i {\hat x}^j {\hat x}^l {\hat x}^m=\frac{4\pi}{15}\left(\delta_{ij}\delta_{lm}-\delta_{il}\delta_{jm}\right)\,.\label{2.82}
\end{equation}
The result is:
\begin{eqnarray}
\underbrace{\left\langle \frac{dE}{dt}\right\rangle}_{(total)}&=& \frac{G}{60}\left\langle \underbrace{ f'_0  \left({ \dddot Q}^{ij}{ \dddot Q}_{ij}\right)}_{GR}
-\underbrace{f''_0 \left({ \ddddot Q}^{ij}{ \ddddot Q}_{ij}\right)}_{f(R)}
 \right\rangle\,.\nonumber\\
\label{2.83}\end{eqnarray}
Precisely, for  $f''_0\rightarrow 0$ and $f'_0\rightarrow \frac{4}{3}$,  Eq.(\ref{2.83}) becomes

\begin{eqnarray}
\underbrace{\left\langle \frac{dE}{dt}\right\rangle}_{(GR)}&=&\frac{G}{45}\left\langle { \dddot Q}^{ij}{ \dddot Q}_{ij} \right\rangle\,.
\label{2.84}\end{eqnarray}which is which is  the  well-known result  of GR \cite{landau,gravitation}. See also \cite{noi-mnras} for the recovering of the correct 
GR-limit.  

An important remark is necessary at this point.  Eq.(\ref{2.83}) can be written as 
\begin{eqnarray}
\underbrace{\left\langle \frac{dE}{dt}\right\rangle}_{(total)}&=& \frac{Gf'_0}{60}\left\langle   \left({ \dddot Q}^{ij}{ \dddot Q}_{ij}\right)
-\frac{1}{m^2}\left({ \ddddot Q}^{ij}{ \ddddot Q}_{ij}\right)
 \right\rangle\,.\nonumber\\
\label{2.83bis}\end{eqnarray}
where the massive mode contribution is evident.  This means that this further term affects both the total energy release as well as the  waveform. This could represent a further signature to investigate such theories in the GW strong-field regime. 

\section{Application to  the binary systems: the  PSR 1913+16 case}
\label{cinque}
Observations coming from PSR 1913+16 can be used to
fix bounds on $f(R)$ parameters. This could be consider a new  test to retain or exclude such theories beside the classical Solar System experiments adopted for GR \cite{gravitation}.
For a binary system, we have to assume that  the motion is Keplerian in the first approximation and  we can average over  orbital periods.
Given  a point mass $m$,  $Q^{ij}(t)$ is 
\begin{eqnarray}
&& Q^{ij}(t)\equiv\int d^3\, {\bf x}\, x^i x^j T^{00}( {\bf x},t)\equiv \nonumber\\ &&
\int\int\int d{\underline x}^1 d {\underline x} ^2 d {\underline x} ^3 \,m\, {\underline x}^i{\underline x}^j \,\times\nonumber\\&&\times {\bf \delta}\left({\underline x}^1-x^1(t)\right){\bf \delta}\left({\underline x}^2-x^2(t)\right){\bf \delta}\left({\underline x}^3-x^3(t)\right)\,,\nonumber\\ \label{2.94}\end{eqnarray}

where ${\underline x}$ is the integration variable and $x$ is the position of the mass \cite{gravitation}. We  define $m$ as the pulsar mass,  $M$ the mass of the companion star, and $\displaystyle{\mu=\frac{GM^3}{(M+m)^2}}$ the reduced mass. This last definition   will be used to account for the fact that $m$ can be small with respect to  $M$. Since the orbit is Keplerian, we can choose  $x^3=0$ being a planar motion. Then  Eq.(\ref{2.94}) reduces to 
\begin{equation}
Q^{ij}(t)=0\,,\qquad \mbox{ for $i$ and/or $j=3$}\,, \label{2.96}
\end{equation}
\begin{eqnarray}
Q^{11}(t)=m(x^1(t))^2\,,\quad Q^{22}(t)=m(x^2(t))^2\,,
\label{2.97}
\end{eqnarray}
\begin{eqnarray}
Q^{12}(t)=Q^{12}(t)=m\left[x^1(t))(x^2(t)\right]\,,\label{2.98}
\end{eqnarray}
where the position in the orbital  plane is  a function of  time \cite{gravitation}. We are going to work in a  parametric representation of the motion \cite{landau, landau1,encounter, hyperbolic} and then let us recast the variables as
\begin{eqnarray}
r&=&a\left(1-\epsilon\cos {\cal E} \right)\,,\quad t=\sqrt{\frac{a^3}{\mu}( {\cal E}-\epsilon\sin  {\cal E})}\,,\nonumber\\
x^1( {\cal E})&=&a\left(\cos  {\cal E}-\epsilon\right)\,,\quad x^2( {\cal E})=a\left(1-\epsilon^2\right)^{\frac{1}{2}}\sin  {\cal E}\,,\nonumber\\
\label{2.99}
\end{eqnarray}
where $r$ is the  orbital radius, $a$, the semi-major axis of the orbit, $\epsilon$, the eccentricity, $ {\cal E}$. the eccentricity anomaly. Over the whole orbit, $ {\cal E}$ ranges from $0$ to $2\pi$. We use $ {\cal E}$, rather than $t$, to locate the body in its orbit, and therefore we have to integrate over $d {\cal E}$

\begin{equation}\label{2.100}
\left\langle f\right\rangle\equiv\frac{1}{T}\int^{T}_{0}dt f(t)\,.
\end{equation}
For a Keplerian orbit $T$ has the value
\begin{equation}
T=2\pi\sqrt{\frac{a^3}{\mu}}\,.
\label{2.101}
\end{equation}
Therefore, if $f(t)=g( {\cal E}(t))$, we may write Eq.(\ref{2.100}) as 
\begin{equation}\label{2.103}
\left\langle f(t)\right\rangle=\frac{1}{2\pi}\int^{\pi}_{0}g( {\cal E})(1-\epsilon\cos  {\cal E})d {\cal E}\,.
\end{equation}
 ${\dddot Q}^{ij}$ can be expressed in terms of the
eccentric anomaly and  then  Eq.(\ref{2.103}) can be used to compute the time average over an  orbital period\footnote{Note  that  we can rise/lower  space indices without regard for sign 
changes because $\eta^{ij}=\delta^{ij}$.}.
We find that  time derivative can be recast as  
\begin{eqnarray}
\frac{d}{dt}=\frac{d {\cal E}}{dt}\frac{d}{d {\cal E}}=\frac{2\pi}{T}(1-\epsilon\cos  {\cal E})^{-1}\frac{d}{d {\cal E}}\,.
\label{2.105}
\end{eqnarray}

From  Eqs.(\ref{2.96})-(\ref{2.98}),  we can write
\begin{eqnarray}
\label{2.111}
{\dddot Q}^{ij}{\dddot Q}_{ij}=\left({\dddot Q}^{11}\right)^2+2\left({\dddot Q}^{12}\right)^2+\left({\dddot Q}^{22}\right)^2\,.
\end{eqnarray}
Let us consider the various orders of derivation.  First, from Eq.(\ref{2.97}) and Eq.(\ref{2.99}) we have 
\begin{eqnarray}
\label{2.112}
Q^{11}({\cal E})=ma^2\left(\cos {\cal E}-\epsilon\right)^2\,.
\end{eqnarray}
Using Eq.(\ref{2.105}) to compute the derivatives, we find
\begin{eqnarray}
{\dot Q}^{11}( {\cal E})=-2ma^2\left(\frac{2\pi}{T}\right) \frac{\sin  {\cal E}\left(\cos  {\cal E}-\epsilon\right)}{1-\epsilon \cos {\cal E}}\,,
\end{eqnarray}
\begin{eqnarray*}
{\ddot Q}^{11}( {\cal E})&=&-2ma^2\left(\frac{2\pi}{T}\right)^2 \left(1-\epsilon \cos {\cal E}\right)^{-3}\times\nonumber\\ &&
\times\left(2\cos^2 {\cal E}-\epsilon\cos  {\cal E}-\epsilon\cos^3  {\cal E}+\epsilon^2-1\right)\,,\nonumber\\
\end{eqnarray*}
and 
\begin{eqnarray}\label{2.113}
{\dddot Q}^{11}( {\cal E})&=&-2ma^2\left(\frac{2\pi}{T}\right)^3 \left(1-\epsilon \cos  {\cal E}\right)^{-5}\sin  {\cal E}\times\nonumber\\ &&
\times\left(\epsilon\cos^2  {\cal E}+2\epsilon^2\cos{\cal E}-4\cos {\cal E} \epsilon^3+4\epsilon\right)\,,\nonumber\\
\end{eqnarray}
 finally
 \begin{eqnarray}\label{2113}
{\ddddot Q}^{11}( {\cal E})&=&-2ma^2\left(\frac{2\pi}{T}\right)^4 \left(1-\epsilon  \cos{\cal E}\right)^{-7}\times\nonumber\\&&\times \left[ \left(16 \epsilon ^3+8 \epsilon ^2+4\right) \cos 2 {\cal E} + (8 \epsilon^2 -3\epsilon )
   \cos {\cal E}+\right.\nonumber\\&&\left.+3 \epsilon  (\cos 3 {\cal E}-4 \epsilon  (2 \epsilon +1))\right]\,.\nonumber\\
\end{eqnarray}
Likewise, from Eq.(\ref{2.97}) and Eq.(\ref{2.99}), we have
\begin{eqnarray}
\label{2.114}
Q^{22} {\cal E}=ma^2\left(1-\epsilon^2\right) \sin^2  {\cal E}\,,
\end{eqnarray}
which leads to the derivatives
\begin{eqnarray*}
{\dot Q}^{22}( {\cal E})=2ma^2\left(\frac{2\pi}{T}\right)\left(1-\epsilon^2\right)\frac{\sin  {\cal E} \cos  {\cal E}}{\left(1-\epsilon \cos  {\cal E}\right)} \,,\nonumber\\
\end{eqnarray*}
\begin{eqnarray*}
{\ddot Q}^{22}( {\cal E})&=&2ma^2\left(\frac{2\pi}{T}\right)^2\frac{\left(1-\epsilon^2\right)}{\left(1-\epsilon \cos  {\cal E}\right)^3}\times\nonumber\\&&\times\left(\cos^2  {\cal E}- \sin^2  {\cal E}-\epsilon  \cos^3 {\cal E}\right)\,,
\end{eqnarray*}
\begin{eqnarray}
{\dddot Q}^{22}( {\cal E})&=&2ma^2\left(\frac{2\pi}{T}\right)^3\frac{\left(1-\epsilon^2\right)}{\left(1-\epsilon \cos  {\cal E}\right)^{5}}\times\nonumber\\&&\times\sin  {\cal E}\left(3\epsilon-4\cos  {\cal E}+\epsilon^2  {\cal E}\right)\,.\label{2.115}
\end{eqnarray}
\begin{eqnarray}
{\ddddot Q}^{22}( {\cal E})&=&2ma^2\left(\frac{2\pi}{T}\right)^4\frac{\left(1-\epsilon^2\right)}{\left(1-\epsilon \cos  {\cal E}\right)^{7}}\times\nonumber\\&&\times
\left[ \left(22 \epsilon ^2-16\right) \cos 2{\cal E} +41 \epsilon  \cos {\cal E}+\right.\nonumber\\&&+\left. (\epsilon^2  (\cos 4 {\cal E}-39)-9\epsilon \cos 3 {\cal E})\right]
\label{2115}
\end{eqnarray}
Finally, from Eq.(\ref{2.98}) and Eq.(\ref{2.99}) we have
\begin{eqnarray}
Q^{12}=Q^{21}=m a^2(1-\epsilon)^{\frac{1}{2}}\sin  {\cal E} (\cos {\cal E}-\epsilon)\,,\nonumber\\
\label{2.116}
\end{eqnarray}
whose  derivatives are
\begin{eqnarray*}
{\dot Q}^{12}&=&ma^2\left(\frac{2\pi}{T}\right)(1-\epsilon)^{\frac{1}{2}}(1-\epsilon\cos  {\cal E})^{-1} \times\nonumber\\&&\times(2\cos^2  {\cal E}-\epsilon\cos  {\cal E}-1)\,,
\end{eqnarray*}
\begin{eqnarray*}
{\ddot Q}^{12}&=&ma^2\left(\frac{2\pi}{T}\right)^2(1-\epsilon)^{\frac{1}{2}}(1-\epsilon\cos {\cal E})^{-3}\sin  {\cal E}\times\nonumber\\&&\times\left(2\epsilon\cos^2  {\cal E}-4\cos  {\cal E}+2\epsilon\right)\,,
\end{eqnarray*}
\begin{eqnarray}
{\dddot Q}^{12}&=&ma^2\left(\frac{2\pi}{T}\right)^3(1-\epsilon)^{\frac{1}{2}}(1-\epsilon\cos  {\cal E})^{-5}\times\nonumber\\&&\times\left(\epsilon^2\cos^2 {\cal E}+3\epsilon\cos  {\cal E}+\epsilon \cos^3  {\cal E}-\right.\nonumber\\&&\left.-3\epsilon^2-4\cos^2  {\cal E}+2\right)\,,
\label{2.117}
\end{eqnarray}
\begin{eqnarray}
{\ddddot Q}^{12}&=&ma^2\left(\frac{2\pi}{T}\right)^4(1-\epsilon)^{\frac{1}{2}}(1-\epsilon\cos  {\cal E})^{-7}\times\nonumber\\&&\times \sin  {\cal E} \left[\left(15 \epsilon ^2+4\right) \cos  {\cal E}+ \left(
   3\epsilon ^3+6\epsilon \right) \cos 2 {\cal E} -\right. \nonumber\\&&\left.+27 \epsilon ^3+\epsilon^2  \cos 3  {\cal E}+18\epsilon \right]
\label{2117}
\end{eqnarray}
When results from Eqs.(\ref{2.113}), (\ref{2.115}), and (\ref{2.117}), together with Eq.(\ref{2.101}) for $T$, are used in Eq.(\ref{2.111}), one finds 
\begin{eqnarray}
{\dddot Q}^{ij}{\dddot Q}_{ij}=4m^2\frac{\mu^3}{a^5}\frac{\left[8\left(1-\epsilon\right)+\epsilon^2\sin^2 {\cal E}\right]}{\left(1-\epsilon\cos  {\cal E}\right)^{6}}\,.\nonumber\\
\label{2.118}
\end{eqnarray}
and
\begin{eqnarray}
{\ddddot Q}^{ij}{\ddddot Q}_{ij}&=&2m^2\frac{\mu^4}{a^8}\frac{1}{\left(1-\epsilon\cos  {\cal E}\right)^{14}}\times\nonumber\\ &&\times 2(\epsilon^2-1)^2\left[41\epsilon \cos  {\cal E}-9-39\epsilon^2+\right.\nonumber\\&&\left.+(22\epsilon^2-16)\cos  2{\cal E}+\epsilon \cos  4{\cal E} \right]^2+\nonumber\\&&+(1-\epsilon)\left[18\epsilon-27\epsilon^3+(4+15\epsilon^2)\cos {\cal E}+\right.\nonumber\\&&\left.+(6\epsilon+3\epsilon^3)\cos2 {\cal E}+\epsilon^2\cos 3 {\cal E}\right]^2\sin^2 {\cal E}+\nonumber\\&& +2\left[(8\epsilon^2-3\epsilon)\cos  {\cal E}+(4+8\epsilon^2+16\epsilon^3)\cos  {\cal E}+\right.\nonumber\\&&\left. 2\epsilon\cos 3  {\cal E}-16\epsilon^3-8\epsilon^2\right]^2
\nonumber\\
\label{2118}
\end{eqnarray}
Substituting  Eq.(\ref{2.118}) into Eq.(\ref{2.103}) and averaging, we have
\begin{eqnarray}
\left\langle {\dddot Q}^{ij}{\dddot Q}_{ij}\right\rangle=\frac{4}{\pi}m^2\frac{\mu^3}{a^5}\int^{\pi}_0 \frac{8\left(1-\epsilon\right)+\epsilon^2\sin^2  {\cal E}}{\left(1-\epsilon\cos  {\cal E}\right)^5}d{\cal E}\,.\nonumber\\
\label{2.119}
\end{eqnarray}
also for Eq. (\ref{2118}) we obtain
\begin{eqnarray}
\left\langle {\ddddot Q}^{ij}{\ddddot Q}_{ij}\right\rangle&=&\frac{m^2}{\pi}\frac{\mu^4}{a^8}\int^{\pi}_0\frac{1}{\left(1-\epsilon\cos  {\cal E}\right)^{13}}\times\nonumber\\&&\times2(\epsilon^2-1)^2\left[41\epsilon \cos  {\cal E}-9-39\epsilon^2+\right.\nonumber\\&&\left.+(22\epsilon^2-16)\cos  2{\cal E}+\epsilon \cos  4{\cal E} \right]^2+\nonumber\\&&+(1-\epsilon)\left[18\epsilon-27\epsilon^3+(4+15\epsilon^2)\cos {\cal E}+\right.\nonumber\\&&\left.+(6\epsilon+3\epsilon^3)\cos2 {\cal E}+\epsilon^2\cos 3 {\cal E}\right]^2\sin^2 {\cal E}+\nonumber\\&& +2\left[(8\epsilon^2-3\epsilon)\cos  {\cal E}+(4+8\epsilon^2+16\epsilon^3)\cos  {\cal E}+\right.\nonumber\\&&\left. 2\epsilon\cos 3  {\cal E}-16\epsilon^3-8\epsilon^2\right]^2 d{\cal E}
\nonumber\\
\label{2119}
\end{eqnarray}
The first term of Eq. (\ref{2.119})  is evaluates using 
\begin{eqnarray*}
\int^\pi_0\frac{d {\cal E}}{(1-\epsilon\cos  {\cal E})^5}=\frac{\pi}{ \left(1-\epsilon^2\right)^{\frac{5}{2}}}P_4\left(\frac{1}{\sqrt{1-\epsilon^2}}\right)\,,
\end{eqnarray*}
where
$\displaystyle{P_4(x)=\frac{1}{8}\left(35x^4-30x^2+3\right)\,.}$
\begin{eqnarray}
\int^\pi_0\frac{d {\cal E}}{(1-\epsilon\cos  {\cal E})^5}=\frac{\pi}{8}\frac{3\epsilon^4+24\epsilon^2+8}{\left(1-\epsilon^2\right)^{\frac{9}{2}}}\,.\nonumber\\
\label{2.120}
\end{eqnarray}
The complete evaluation of Eq. (\ref{2.119}) is 
\begin{eqnarray}
\left\langle {\dddot Q}^{ij}{\dddot Q}_{ij}\right\rangle=\frac{1}{2}m^2\frac{\mu^3}{a^5}\frac{  25\epsilon^4+196\epsilon^2+64 }{\left(1-\epsilon^2\right)^{\frac{7}{2}}}\,.\nonumber\\
\label{2.121}
\end{eqnarray}
and for the Eq. (\ref{2119}) we do not have an analytical solution of the integral but, only a numerical result that will be insert  in the following equations.

The above results  apply for the motion of a body of mass $m$ in a Keplerian orbit about a second body of mass $M$. Therefore, we can evaluate the overall loss rate due to the motion of both bodies.
Let the subscript $1$ denote the position of the pulsar $m$ and $2$ that of the companion $M$ and, as above, let the coordinate origin be at the barycenter \cite{smart,roy}. This condition gives
\begin{eqnarray}
mx_1^i+Mx^i_2=0\,,
\end{eqnarray}
and then
\begin{eqnarray}
x^i_2=-\frac{m}{M}x_1^i\,.
\label{2.122}
\end{eqnarray}
The averall momentum $Q^{ij}$ for the system consisting of both $m$ and $M$ is

\begin{eqnarray}
Q^{ij}=mx^i_1x^j_1+Mx^i_2x^j_2=\frac{m}{M}(m+M)x^i_1x^j_1\,,
\label{2.123}
\end{eqnarray}
where we have used Eq.(\ref{2.122}) to express the momentum in terms of the motion of $m$. 
 Averaging for the binary system, we obtain
\begin{eqnarray}
\left\langle {\dddot Q}^{ij}{\dddot Q}_{ij}\right\rangle=\frac{G^3m^2M^7}{2a^5(m+M)^4}\frac{25\epsilon^4+196\epsilon^2+64}{(1-\epsilon^2)^\frac{7}{2}}\,.\nonumber\\
\label{2.125}
\end{eqnarray}
We do not measure $\displaystyle{\frac{d {\cal E}}{dt}}$ directly but  the change of orbital period $T$ induced by $\displaystyle{\frac{d {\cal E}}{dt}}$. 
To this end, we remember that the semi-major axis of the orbit is \cite{landau,landau1,smart,roy,dixon}
\begin{eqnarray}
a'=\frac{m+M}{M}a\,,
\end{eqnarray}
where it has to be recalled that $a$ is the semi-major axis of the pulsar orbit. The total energy $E$ of a Keplerian binary system is then

\begin{eqnarray}
E=-\frac{GmM}{2a'}=-\frac{GmM^2}{2a(m+M)}\,,
\label{2.127}
\end{eqnarray}
from which
\begin{eqnarray}
a=-\frac{GmM^2}{2(m+M)}\frac{1}{E}\,.
\label{2.128}
\end{eqnarray}
The orbital period $T$ can be related to the energy $E$ by combining Eq.(\ref{2.101}) and Eq.(\ref{2.128}). 
The result is
\begin{eqnarray}
T=-\pi GE^{\frac{3}{2}}\left(\frac{m^3M^3}{2(m+M)}\right)^{\frac{1}{2}}\,.
\label{2.129}
\end{eqnarray}
By taking the time derivative and Eq.(\ref{2.127}) to restore the parameter $a$, we find that
\begin{eqnarray}
\frac{dT}{dt}={\dot T}=6\pi\frac{(m+M)^2}{m}\sqrt{\frac{a^5}{G^3M^7}}\frac{dE}{dt}\,.
\label{2.130}
\end{eqnarray}
Let us now use the published numerical values for the specific example of PSR 1913+16 to numerically
evaluate the above equations . The results will be included into  Eq.(\ref{2.83}) to evaluate $\displaystyle{\left\langle \frac{dE}{dt}\right\rangle}$ from which $ \displaystyle{\frac{dT}{dt}}$ can be estimated using  Eq.(\ref{2.130}).
We use the values from Taylor et al. \cite{hulse,taylor} for PSR 1913+16 reported  in Table I.

\begin{table}[htbp]
\begin{center}
\begin{tabular}{|l|c|r|}
\hline
 \bfseries PSR\, 1913+16 & chacteristic features \\
\hline
\hline
pulsar mass & $m =1.39 M_{\odot}$\\
\hline
 companion mass & $M = 1.44M_{\odot}$ \\ 
\hline
inclination angle &$\sin i = 0.81$\\
\hline
orbit semimajor axis  &$a =  8.67\times10^{10}cm$\\
\hline
eccentricity &$\epsilon = 0.617155$\\
\hline
gravitational constant &$G = 6.67\times10^{-8}dyn\, cm^2\,g^{-2}$\\
\hline
speed of light &$c = 2.99\times 10^{10} cm\, sec^{-1}$\\
\hline
\end{tabular}
\end{center}
\label{uno}
\caption{Values from Taylor et al. for PSR 1913+16 \cite{hulse,taylor}.}
\end{table}

First we find, from  Eq.(\ref{2.130}), that for PSR 1913+16
\begin{equation}
{\dot T}=2.21\times10^{-44}\left(\frac{sec^3}{g\,cm^2}\right)\frac{dE}{dt}\,.
\label{2.131}
\end{equation}
\begin{equation}
\left\langle \left({\dddot Q}^{ij}\right)^2\right\rangle=2.78\times10^{92}\left(\frac{g\,cm^2}{sec^3}\right)^2\,,\label{2.133}
\end{equation}
\begin{equation}
\left\langle \left({\ddddot Q}^{ij}\right)^2\right\rangle=1.29\times10^{96}\left(\frac{g\,cm^2}{sec^3}\right)^2\,
\label{2133}
\end{equation}
 Then, the averaged total radiation rate for the binary system in  $f(R)$- gravity is found
from Eq.(\ref{2.83}), that is \footnote{Note that for dimensional reasons we have to restore the factor of $c^5$ to the denominator of $\frac{dE}{dt}$.}
\begin{eqnarray}
\left\langle \frac{dE}{dt}\right\rangle&=& \left[1.32\times 10^{31} f'_0-6.10\times 10^{34} f''_0\right] \left(\frac{g\, cm^2}{sec^3}\right)\,.\nonumber\\ \label{2.134}
\end{eqnarray}

Using this value in Eq.(\ref{2.131}), we find
\begin{eqnarray}
{\dot T}_{f(R)}&=&2.92\times10^{-13}f'_0-1.34\times10^{-9}f''_0 \left(\frac{sec}{sec}\right)\,,\nonumber\\\label{2.135}
\end{eqnarray}
as we can see from the above equation the orbital period depends strongly on the choice theory. Now, given the value of $f'_0$, ({\it i.e. }$\displaystyle{f'_0=\frac{4}{3}}$), we determine the value of $f''_0$ that falls within the limits observed by Hulse and Taylor \cite{hulse,taylor}.
We remember that they predicted an upper and lower limit in the observation of the orbital period that is about $3.8\times 10^{-12}\leq{\dot T}\leq2.6\times 10^{-12} $ and the limit for GR is
\begin{equation}
{\dot T}_{GR}\simeq3.36\times10^{-12}\left(\frac{sec}{sec}\right)\,.\label{2.136}
\end{equation}
We immediately recover GR limit putting $\displaystyle{f'_0=\frac{4}{3}}$ and $f''_0=0$.  
In Fig. 1 is shown a plot of $\dot T$ from (\ref{2.135}) for PSR 1913+16 as a function of $f''_0$ parameter. 
\begin{figure}[ht!]
\label{figure1}
\leavevmode
\includegraphics[scale=0.5]{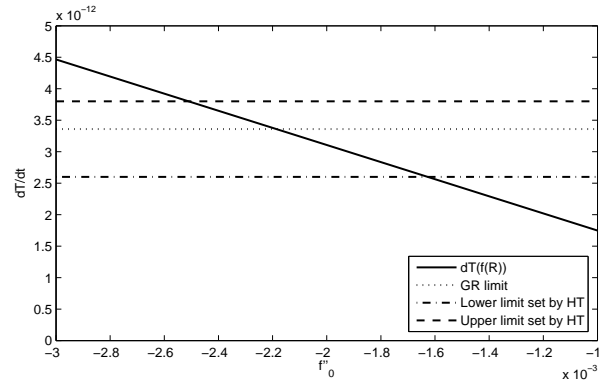}
\caption{Orbital decay rate for PSR 1913+16 in $f(R)$-gravity. Upper limit set by Taylor  et al. in dashed line. 
GR limit $3.36\times 10^{-12}$ in dotted line and the lower limit set by Taylor et al. in dashdot line. While in solid line is plotted ${\dot T}_{f(R)}$. }
\end{figure}
In Fig. 1,  the observational limits on $\dot T$ are  indicated together with  the GR limiting value (\ref{2.136}). The range  $-2.63\times 10^{-3}\leq f''_0 \leq -2.25\times 10^{-3}$  well fits with these  observational  limits \cite{hulse,taylor}.
In other words, we can conclude that $f(R)$-gravity is not excluded by the Hulse and Taylor  observations on binary  pulsar. On the other hand,  such  observations contribute to fix the range of viability of such theories.

\section{Concluding remarks}
\label{sei}
 In this paper, we developed the post-Minkowskian limit of analytic $f(R)$-gravity models  in the Jordan frame to calculate the gravitational radiation emitted by a binary system. One of  the  results   is that
 the  quadrupole-radiation, in $f(R)$-gravity and in GR,  occurs independently of the detailed
internal structure of the stellar bodies. It depends  on the masses of the two bodies, on the orbital parameters and on the details of the gravitational theory. 
Further massive modes  emerge and they are directly related to the analytic parameters of  $f(R)$-gravity, that is the coefficients  $f'_0$ and $f''_0$ of the Taylor expansion.  This fact is  relevant since it does not depend on specific $f(R)$-models  but it  is a general feature.

As a consequence,   the
theoretical quadrupole radiation rate, calculated according to the theory,  can be confronted to  
 binary system observations to fix the parameters of the theory.  Specifically,  the radiation rate is a function of $f'_0$ and $f''_0$.
As we can see from Fig. 1 or, equivalently from Eq.  (\ref{2.135}), the predicted range of the
time derivative of  the orbital period for  PSR 1913+16
 is compatible with the observational uncertainty established by Hulse and Taylor.
\cite{grgOdi}. This means that observations can fix the parameters of the theory.
These results pose interesting problems related to the strict validity of GR. It seems that it works very well at local scales (Solar System) where effects of further gravitational degrees of freedom cannot be detected. As soon as one is investigating larger scales, as  those of  galaxies,  clusters of galaxies, etc.,  further corrections can be introduced in order to explain both astrophysical large-scale dynamics \cite{noi-mnras} and cosmic evolution \cite{f(R)-noi,pogosian}.  Alternatively, huge amounts of dark matter and dark energy have to be invoked to explain the phenomenology, but, up today there are no final evidences for these new constituents at fundamental level. What we have shown is that the Hulse and Taylor experiment, beside confirming GR, does not exclude Extended Theories of Gravity \cite{book} including GR as a particular case.

 Furthermore, the fact that, up to now, only massless gravitational waves have been investigated could be a shortcoming preventing the possibility to find out other forms of gravitational waves. Tests in this sense could come out, for example, 
 from the stochastic background of gravitational waves  where massive modes could play a crucial role in 
 the cosmic background spectrume \cite{bellucci,felix}.

\section*{ACKNOWLEDGEMENTS}
We wish to thank L. Milano for  discussions and  comments on the topics of  this paper.


\end{document}